\documentclass[pra,preprint,showpacs]{revtex4-1}

\usepackage{amsmath,latexsym,amssymb,amsfonts}
\usepackage[usenames,dvipsnames]{color}
\usepackage[colorlinks]{hyperref}

\usepackage{graphicx}
\usepackage[applemac]{inputenc}
\usepackage[T1]{fontenc} 

\newcommand{\beq}{\begin{equation}}
\newcommand{\eeq}{\end{equation}}
\newcommand{\bea}{\begin{eqnarray}}
\newcommand{\eea}{\end{eqnarray}}
\def\haf#1{{{#1}\over 2}}

\newcommand{\eps}{\epsilon}

\newcommand\ltwid{\mathrel{
 \raise.3ex\hbox{$<$\kern-.75em\lower1ex\hbox{$\sim$}}}}

\begin{document}

\preprint{UdeM-GPP-TH-15-242}
\preprint{DAMTP-2015-30}

\title{Tunneling decay of  false domain walls: the silence of the lambs}
\author{Mareike Haberichter$^{b,d}$}
\email{M.Haberichter@kent.ac.uk}
\author{Richard MacKenzie$^{a}$}
\email{richard.mackenzie@umontreal.ca}
\author{M.~B.~Paranjape$^{a,b,c}$}
\email{paranj@lps.umontreal.ca}
\author{Yvan Ung$^{a}$\footnote{address after Sept. 1st, 2015: School of Physics and Astronomy,
University of Minnesota, 116 Church St SE, Minneapolis, MN 55455 USA}}
\email{klingon_ecology@hotmail.com}
\affiliation{$^a$Groupe de Physique des Particules, D\'epartement de physique, Universit\'{e} de Montr\'{e}al, C.~P.~6128, Succursale Centre-ville, Montreal, Qu\'{e}bec, Canada, H3C 3J7}
\affiliation{$^b$Department of Applied Mathematics and Theoretical Physics, University of Cambridge, Wilberforce Road, Cambridge, CB3 0WA UK}
\affiliation{$^c$St. John's College, University of Cambridge, Cambridge, CB2 1TP UK}
\affiliation{$^d$School of Mathematics, Statistics and Actuarial Science, University of Kent, Canterbury, CT2 7NF UK}

\begin{abstract}
We study the decay of ``false'' domain walls, which are metastable states of the quantum theory where the true vacuum is trapped inside the wall, with the false vacuum outside.  We consider a theory with two scalar fields, a shepherd field and a field of sheep.  The shepherd field serves to herd the solitons of the sheep field so that they are nicely bunched together.  However, quantum tunnelling of the shepherd field releases the sheep to spread out uncontrollably.  We show how to calculate the tunnelling amplitude for such a disintegration. 
\end{abstract}

\maketitle

\newpage

\section{Introduction \label{sec-1}}
The notion of a vacuum in field theory is quite a bit richer than that in everyday language. In a theory with scalar fields, any state that is a local minimum of the potential energy density is a vacuum. More than one vacuum can exist, and not all vacua need have the same energy. The lowest such state is called the true vacuum; any higher-energy local minimum is a false vacuum. 

The most common reason for a multiplicity of vacua is spontaneous symmetry breaking. Breaking of a discrete symmetry gives rise to discrete vacua whereas breaking of a continuous symmetry gives rise to a continuum of vacua. In this paper we restrict ourselves to discrete vacua. Depending on the model, a number (possibly infinite) of vacua of any energy can exist. Field configurations that interpolate between different degenerate vacua, true or false, give rise to domain walls (see for example \cite{ms}), which themselves are stable or metastable.  Domain walls are very important in cosmology \cite{v}, and in condensed matter physics \cite{cm}, where they quite often arise in the study of first order phase transitions.  A full review of domain walls and their applications is beyond the scope of this article, and we refer the reader to the cited literature. It is clear that domain walls are very important in wide swaths of theoretical physics.

In this paper we consider the special case of domain walls interpolating between distinct false vacua, with true vacuum in the core of the domain wall. The very existence of domain walls in our model is due to a subtle interplay between the energetics of two fields.  The domain wall we consider entraps several quanta of a  quantum field, $\phi$, which we call a field of sheep, which on their own would separate to infinity, leaving behind regions of true vacuum.  It is the sheep field that is in the false vacuum outside the domain wall, but in regions passing through the true vacuum inside the domain wall.  These sheep are herded into staying together, inside the domain wall, by a field we call the shepherd field, $\psi$, which is in its true vacuum outside the domain wall, but in fact in its false vacuum inside the domain wall.  The shepherd field is unstable to quantum tunnelling to its true vacuum; once this occurs, the sheep are without a shepherd and will spread out to infinity.  We compute the amplitude for such a decay in a specific model analytically, aided by numerical calculations, within a well-defined approximation scheme.  We have also studied an analogous model \cite{am}, but in that model, there are not multiple quanta of one field trapped inside the domain wall.
\section{The model}
Consider the model defined by the Lagrangian density
\beq
{\cal L}= {\frac{1}{2}}{\left(\partial_\mu\psi\partial^\mu\psi +\partial_\mu\phi\partial^\mu\phi\right)} -V(\psi,\phi)
\eeq
with
\beq
V(\psi,\phi)=V_\psi(\psi)+V_\phi(\phi)+V_{\psi\phi}(\psi,\phi)-V_0
\eeq
where
\beq
V_\psi(\psi)=\alpha (\psi+a)^2((\psi-a)^2+\epsilon_\psi^2))\label{Vpsi}
\eeq
\beq
 V_\phi(\phi)=\beta(\sin^2(\pi\phi)+\epsilon_\phi\sin^2(\pi\phi/2))\label{Vphi}
\eeq
while
\beq
V_{\psi\phi}(\psi,\phi)=\lambda\frac{(\psi-a)^2((\psi+a)^2+\epsilon_\psi^2))}{(V_\phi(\phi)-V_\phi(1/2)^2+\gamma^2}
\eeq
$\lambda$ is considered to be very small compared to $\alpha$ and $\gamma$ is considered small compared to $V_\phi(1/2)$.   We impose that  $V(\psi,\phi)\ge 0$. The first three terms are positive semi-definite, and then $V_0$ is chosen so that $V(\psi,\phi)$ vanishes at its global minimum.  The $\phi$ field is essentially a sine-Gordon field (with alternating true and false vacua) while the $\psi$ field has a rather standard double well potential with a small asymmetry.  Both fields admit kink type solitons when $\epsilon_\phi=\epsilon_\psi=0$.  The interaction between the two fields is constructed to disallow the passage of the $\phi$ solitons through the $\psi$ solitons.  The specific potential that we use is not important; the only feature that is crucial is this ``non-commutativity'' of solitons, and any potential that achieves the result is adequate.

\begin{figure}[ht]
\begin{center}
\includegraphics[width=3.in]{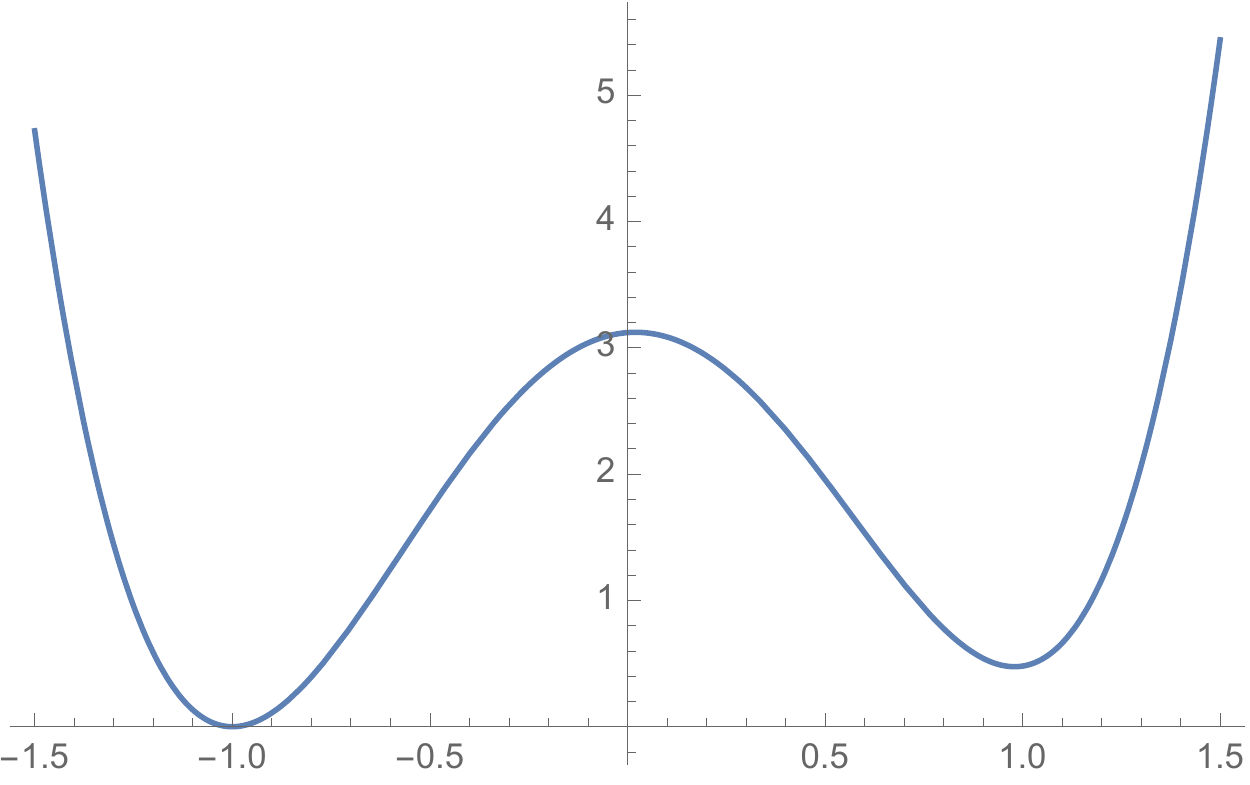}
\end{center}
\caption{(Color online) The potential  $V_\psi$ for the field $\psi$ with $a=1$.
}
\label{fig1}
\end{figure}

\begin{figure}[ht]
\begin{center}
\includegraphics[width=3.in]{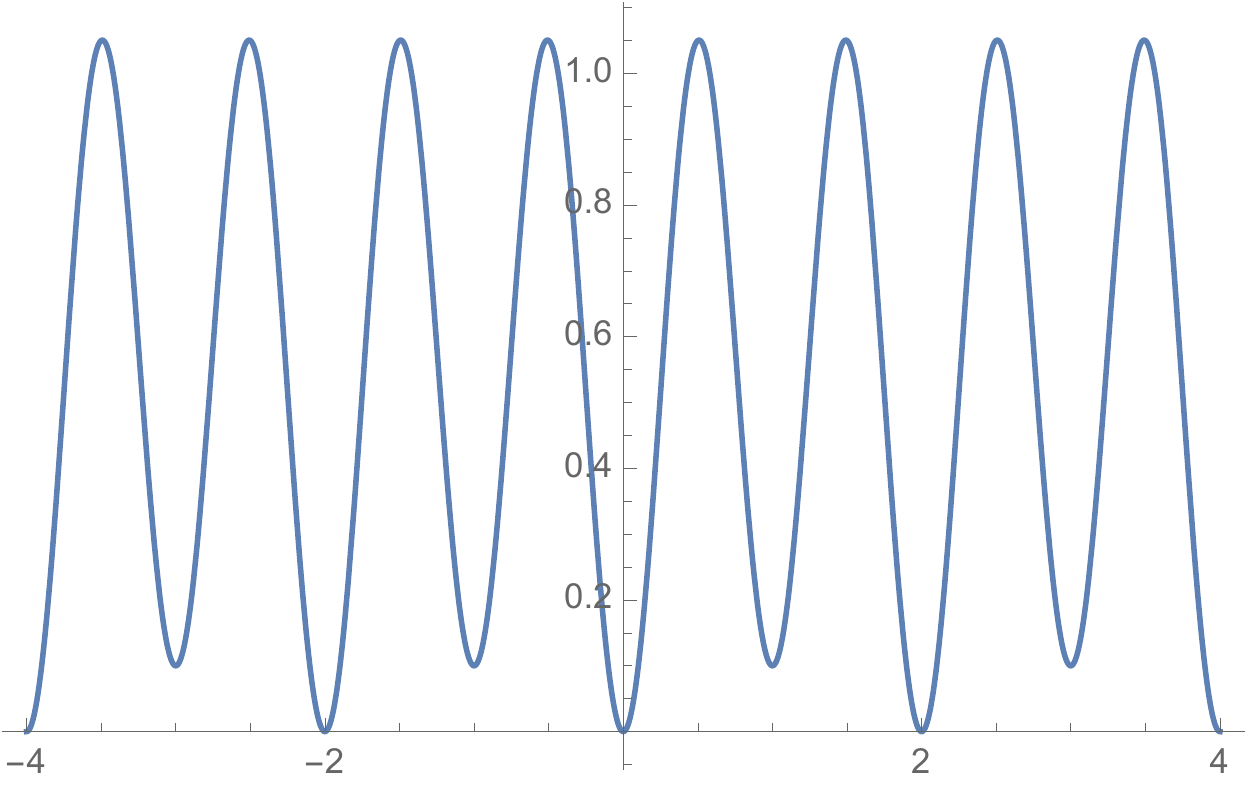}
\end{center}
\caption{(Color online) The potential  $V_\phi$ for the field $\phi$.
}
\label{fig2}
\end{figure}

It is easy to see that the global minimum of $V_\psi$ occurs at $\psi=-a$ while a local minimum occurs near $\psi=a$, as can be seen in Fig.\eqref{fig1}.  These values are mildly perturbed by $V_{\psi\phi}(\psi,\phi)$ but for small values of $\lambda$; however the changes can be made arbitrarily small.  The minimum of $V_\phi(\phi)$ occurs at even integers $\phi=2k$ for any integer $k$,  with a slightly higher local minimum occurring at odd integers $\phi=2k+1$, and local maxima occur at each half odd integer, as can be seen in Fig.\eqref{fig2}.  Again, these values are mildly perturbed by $V_{\psi\phi}(\psi,\phi)$ but for small values of $\lambda$ the changes can be made arbitrarily small.  

Obviously for $\lambda=0$, the true vacuum corresponds to $\{\psi=-a, \phi=2k\}$ while $\{\psi=- a, \phi=2k+1\}$ or $\{\psi=a, \phi=2k$ or $2k+1\}$ correspond to the false vacuum.  Of course with $\lambda$ non-zero, the values of the fields at the minima will be slightly modified, however the structure of the false and true vacua will remain unchanged.  In the ensuing discussion we will identify the vacua by the values that the fields take when $\lambda=0$, but always with the understanding that the actual values that the fields take are slightly perturbed.  

\begin{figure}[ht]
\begin{center}
\includegraphics[width=5in]{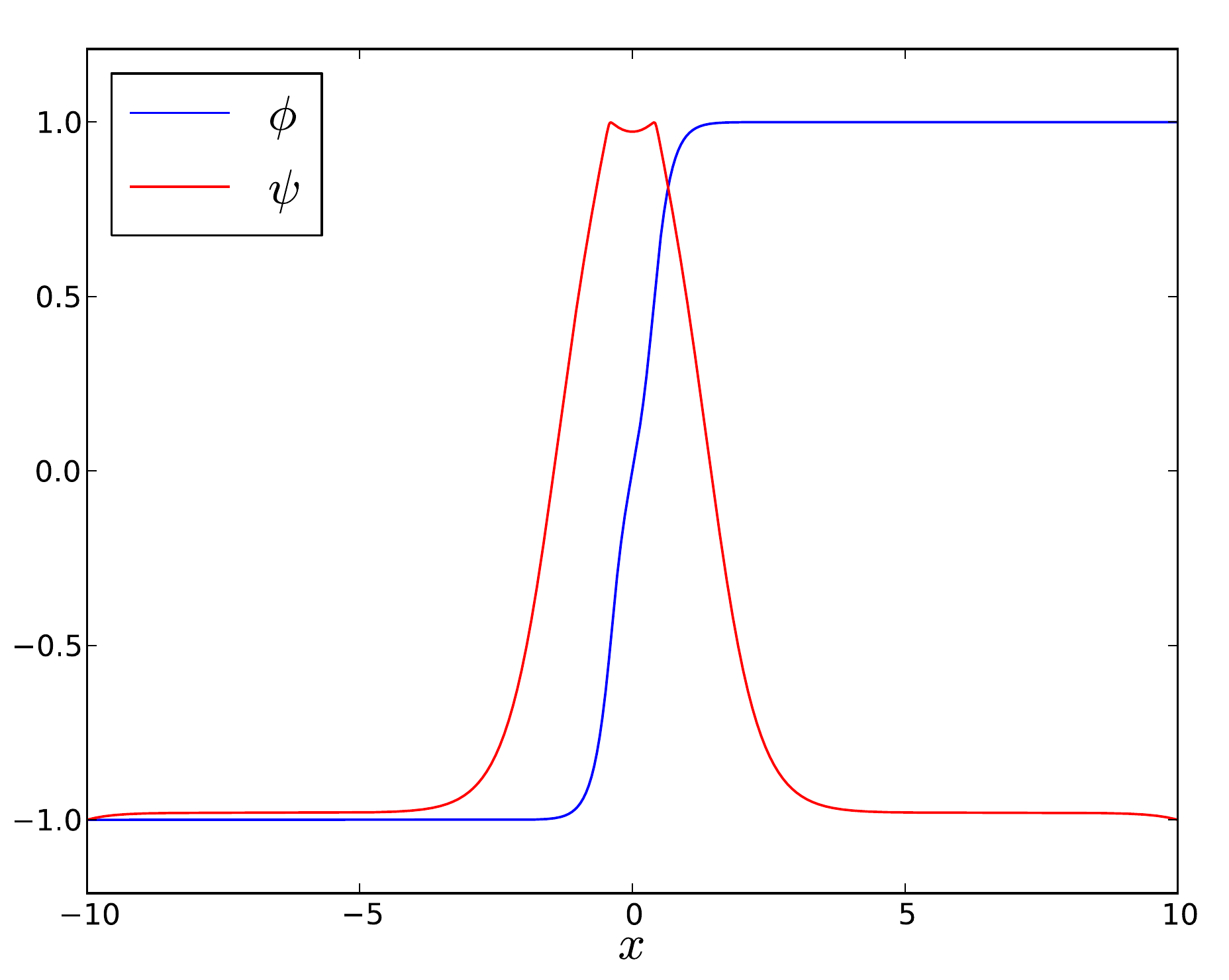}
\end{center}
\caption{(Color online) The metastable domain wall configuration, $\phi$ passing from $-1\to 1$ and $\psi$ passing from $-1\to 1\to -1$.  For this and all subsequent figures, the parameters were chosen as $a=1, \eps_\phi=0.01, \eps_\psi=0.5, \lambda=0.1, \alpha=0.5, \beta=0.5,\gamma=0.01$
}
\label{fig3}
\end{figure}

As there are  several discrete vacua, we can easily imagine configurations corresponding to domain walls.  Consider a domain wall interpolating between the two false vacua, $\phi= 2k+1$ and  $\phi= 2k'+1$ but $\psi=-a$, its true vacuum.  Thus $\phi$ interpolates from $2k+1$ to $2k'+1$ while the field $\psi=-a$ remains at its true minimum.  Such a domain wall will be immediately unstable to separating into, in principle, several subsidiary domain walls.   Consider the specific example where the $\phi$ field interpolates from $\phi=-1\to\phi=1$.  Such a domain wall will be unstable, separating into two half domain walls that interpolate from $\phi=-1\to \phi=0$ and then from $\phi=0 \to \phi=1$ while the $\psi$ field remains always at $\psi=-a$.  The intermediate region being true vacuum, $\phi=0,\psi=-a$, the two half domain walls feel a repulsive force and will accelerate away from each other leaving behind them true vacuum.  However,  taking into account the interaction term, such a false domain wall can be rendered metastable (classically stable, but unstable to quantum tunnelling) by imposing some transitions in the $\psi$ field to its false vacuum.   The $\psi$ field must makes its transition to the left of the point where $\phi$ field begins its transitions, and $\psi$ must return back to its true vacuum only after $\phi$ has completed its transitions and has reached 1.  There would be a soliton and an anti-soliton pair in the $\psi$ field.   The configuration corresponds to something like Fig.\eqref{fig3}, which is a numerical solution of the equations of motion, and has been obtained using the gradient flow method \cite{gfm}.  

In the stability analysis,  the interaction term makes the following crucial contribution. The point is that if $\phi$ passes through the value $1/2$ when $\psi$ is such that the numerator in the interaction term is not very small or even zero (this means that $\psi\ne\pm a$), then the denominator of the interaction term becomes $\gamma^2$, which is chosen to be very small, and thus the interaction term can be made to blow up.  In this way,  the contribution to the energy of the interaction term can be made arbitrarily large, which we assume, and therefore such a situation will be energetically disfavoured.  What this means specifically is that the $\phi$ solitons may not pass through the $\psi$ solitons (in which necessarily occur regions where the $\psi$ field is significantly different for $\psi=\pm a$) without the energy blowing up.  There is an energy barrier to such a process.  Thus if $\psi$ is forced to make a transition from $-a$ to $a$ only before or after $\phi$ makes its transitions, the solitons of the $\phi$ field cannot pass through those of the $\psi$ field without encountering an insurmountable energy barrier.   But then the solitons of the $\phi$ field, the sheep, are trapped within the solitons of the $\psi$ field, the shepherd.  

Inside the domain wall, two adjacent $\phi$ half-solitons (say, one going from $\phi=2k$ to $2k+1$ and the other from $2k+1$ to $2k+2$) attract each other.  This is because the energy is minimized by reducing the region over which $\phi$ is in its false vacuum, $\phi=2k+1$.  Thus the net $\phi$ soliton passes from even integer to even integer.  But then these full solitons  further minimize their energy by separating as far as possible from each other, as this minimizes any gradient energies that occur due to their proximity; hence, they have a repulsive interaction.  The half-solitons on either end are also simply repelled by the adjacent full solitons.  

\begin{figure}[ht]
\begin{center}
\includegraphics[width=5in]{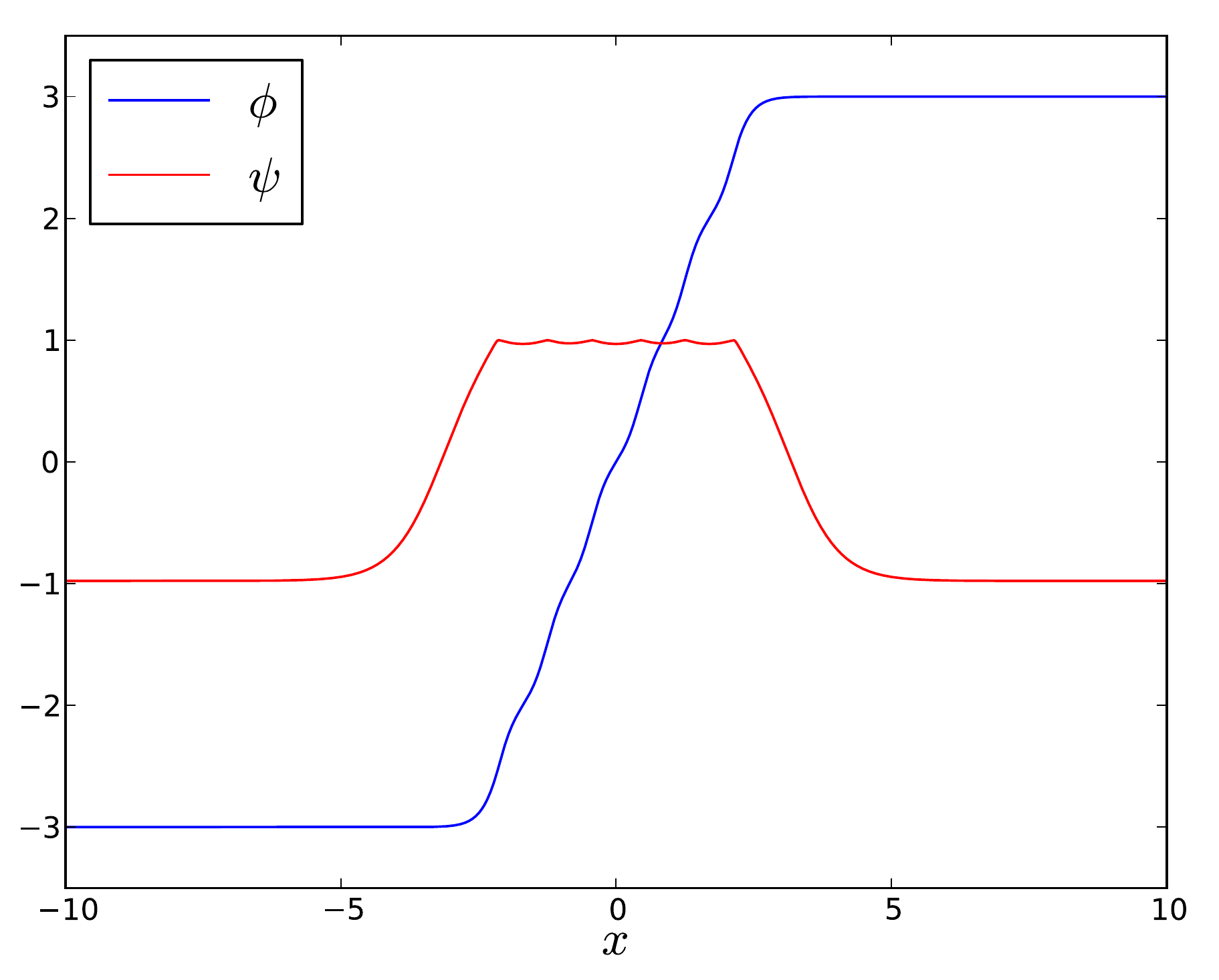}
\end{center}
\caption{(Color online) The metastable domain wall configuration, $\phi$ passing from $-3\to 3$ and $\psi$ passing from $-1\to 1\to -1$.
}
\label{fig4}
\end{figure}
\begin{figure}[ht]
\begin{center}
\includegraphics[width=5in]{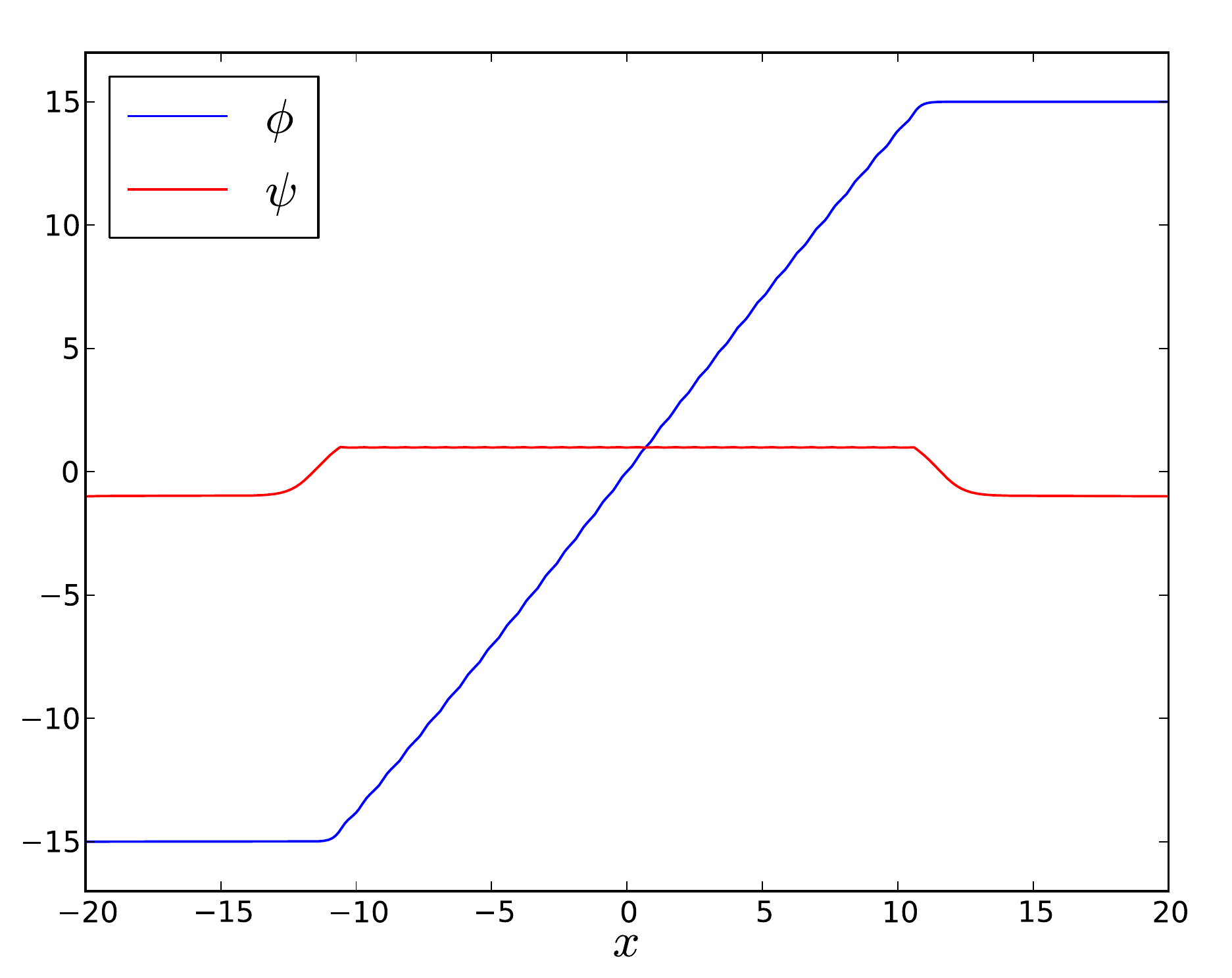}
\end{center}
\caption{(Color online) The metastable domain wall configuration with 31 sheep, $\phi$ passing from $-15\to 15$ and $\psi$ passing from $-1\to 1\to -1$.
}
\label{fig5}

\end{figure}
\begin{figure}[ht]
\begin{center}
\includegraphics[width=7cm]{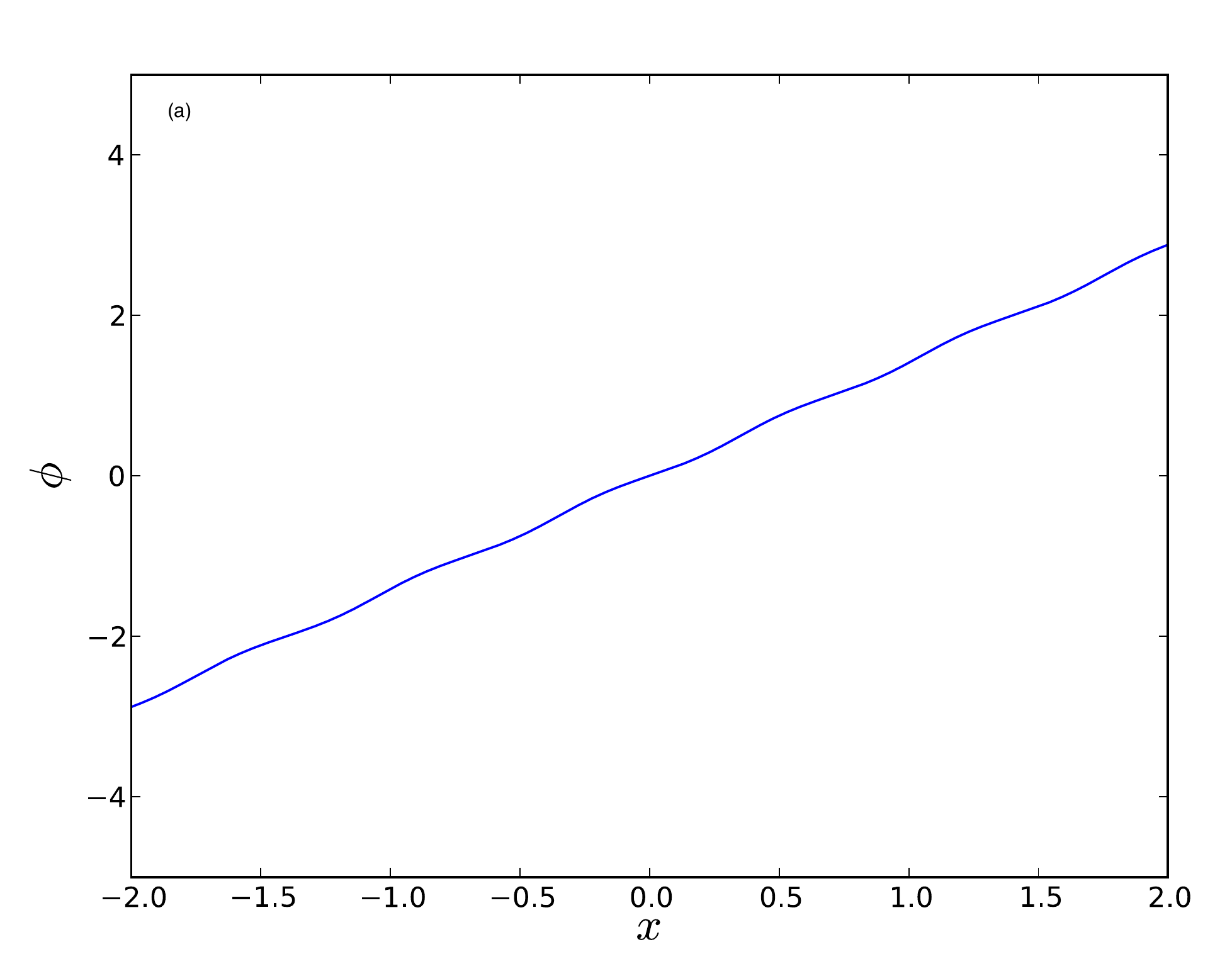}\qquad
\includegraphics[width=7cm]{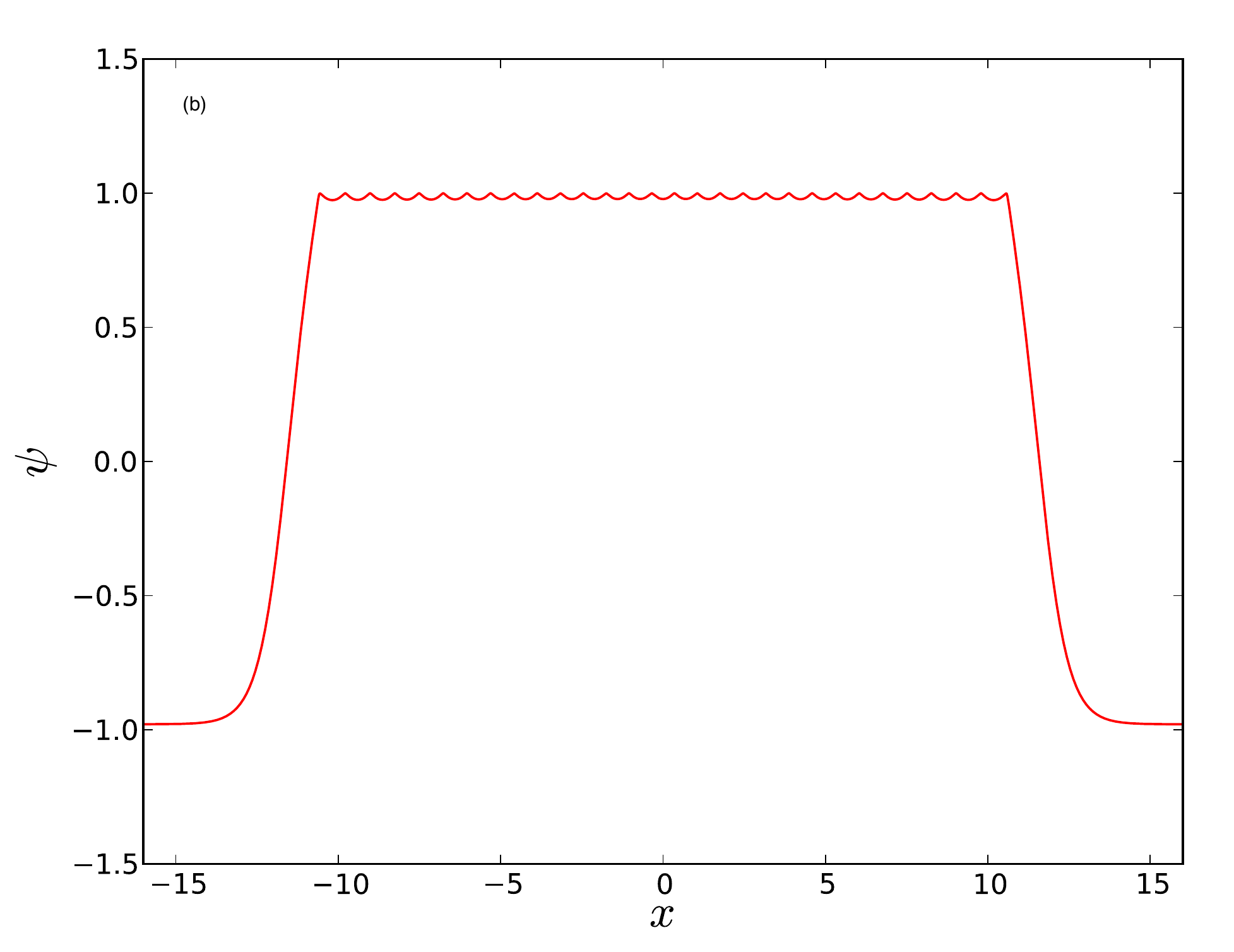}
\end{center}
\caption{(Color online) (a) Zoom of the metastable domain wall configuration with 31 sheep for the field $\phi$.  (b)  Zoom of the metastable domain wall configuration with 31 sheep for the field $\psi$.
}
\label{fig5z}
\end{figure}

Thus in the example shown in Figure \eqref{fig4}, in which $\phi$ interpolates from -3 to 3, the half-soliton from -3 to -2 is repelled by the full solitons from -2 to 0, which also repels the subsequent full soliton from 0 to 2 which in turn repels the final half-soliton from 2 to 3.  The $\psi$ (shepherd) solitons have the false vacuum $\psi\approx a$ in between them, so they attract one another.   The sheep want to separate as far as possible from one another, but they cannot do so without pushing the shepherd solitons further and further apart.  The energy in doing so, increases linearly with the separation of the shepherd solitons.  The competing energy decrease coming from the separating sheep is at best decreased to a constant value (the total mass of sheep) as the separation between the sheep becomes large.  Hence, energetically, the shepherd solitons force the sheep solitons to be herded together at a finite size where the outward pressure from the sheep is balanced against the inward vacuum pressure pushing the shepherds together.  The whole system forms a classically stable configuration.  An even more extreme example is given in Figure \eqref{fig5} and the corresponding zooms in Figures \eqref{fig5z}, which contains 31 sheep.


However, quantum mechanically, the herded sheep are actually in a metastable configuration.  Tunnelling processes can permit the $\psi$ field to make a quantum transition to its true vacuum at all points where the $\phi$ field is near its true vacuum (and specifically away from values where $V(\phi)=V(1/2)$).  Such transitions truly liberate the sheep, which can now move apart freely.    
\section{Analytical calculation of the decay rate in the free kink approximation}
As we have just noted, the soliton is metastable: it will decay via quantum tunnelling.  The tunnelling transition is mediated by an instanton, which is called a bounce.  The bounce corresponds to a trajectory in Euclidean time, $t\to -i\tau$.  This changes the sign of the kinetic energy, $T\to-T$, and therefore bounce configuration can be thought of as motion in the potential $-V$ where we include the spatial gradient terms as part of the potential.  The Euclidean action can be written as $S_E=T+V$ and the conserved Euclidean ``energy'' is $E=T-V$.  The bounce begins at the metastable configuration, arrives at the configuration which will materialize upon the event of the quantum transition, and then ``bounces back'', that is, returns to the initial configuration.  Therefore, the bounce will begin at the metastable soliton as in Fig.(\ref{fig6}a) followed by an evolution during which the field $\psi$ which is initially equal to $a$ in the region in between the two half-solitons sinks down until it finally makes it to $\psi=-a$, corresponding to a configuration of two free half-solitons as depicted in Fig.(\ref{fig6}b).  This configuration would actually correspond to the unstable configuration at the top of the potential barrier.  The two half-solitons must be created, however,  with some kinetic energy, since during the Euclidean evolution, the ``energy'' $T-V$ is conserved, and $-V$ is more negative at this configuration of two essentially free solitons.  In terms of the Minkowski dynamics, since this configuration corresponds to the top of the barrier, it has much more potential energy, $+V$,  than the initial configuration.  Then in the Euclidean dynamics, it has potential energy $-V$; hence there has to be compensating kinetic energy.  The solitons then separate and rise up the (negative) potential until they stop, after which they bounce back.  When they reach the bounce point, that is the configuration at which they will materialize when the quantum tunnelling transition occurs, and from that point on Minkowski evolution takes over.  

The Euclidean Lagrangian is given by
\beq
L_E=\int dx \left(\haf 1 \dot\phi^2+\haf 1 \dot\psi^2+\haf 1 \phi'^2+\haf 1 \psi'^2 +V(\phi,\psi)\right).
\eeq
The fields depend on time only through their dependence on the positions of the solitons.  Indeed, in a first approximation, the fields depend independently on  the positions of the solitons, which is the appropriate collective coordinate describing their evolution.  This means we can approximate $\phi(x,x_1,x_2)=\phi_1(x-x_1) +\phi_2(x-x_2)$ while $\psi(x,x_1,x_2)=\psi_1(x-x_1) +\psi_2(x-x_2)+\psi_0$ where $\psi_0$ is a necessary constant offset.  This approximation is certainly valid when the solitons are separated from one another.  It is not valid when they are very close, and we will have to resort to numerical calculations to find the action there, but we leave that for later.    For example, for $\dot\psi$ we get
\bea
\dot\psi&=&\partial_{x_1}\psi (x,x_1,x_2)\dot x_1+\partial_{x_2}\psi (x,x_1,x_2)\dot x_2\cr
&=&\partial_{x_1}\psi_1 (x-x_1)\dot x_1+\partial_{x_2}\psi_2 (x-x_2)\dot x_2\cr
&=&-\partial_{x}\psi_1 (x-x_1)\dot x_1-\partial_{x}\psi_2 (x-x_2)\dot x_2.
\eea
$\partial_x\psi_1(x-x_1)=\psi_1'$ is a function that is sharply peaked around $x_1$ while  $\partial_x\psi_2(x-x_2)=\psi_2'$ is sharply peaked around $x_2$; thus, the product of these functions essentially vanishes.  This gives
\beq
L_E=\int dx \left(\haf 1 \phi_1'^2\dot x_1^2+\haf 1 \phi_2'^2\dot x_2^2+\haf 1 \psi_1'^2\dot x_1^2+\haf 1 \psi_2'^2\dot x_2^2+\haf 1 \left(\phi_1'^2+\phi_2'^2+\psi_1'^2+\psi_2'^2\right) +V(\phi,\psi)\right).
\eeq
Furthermore, the nature of the solitons at $x_1$ and $x_2$ in each field are essentially identical: they are simply the ``kink''-type solitons of an (almost) doubly degenerate (actually multiply degenerate, but only adjacent minima are relevant here) potential.  Therefore we will compute the contribution of the kinks in the limit that the potential is exactly degenerate, and we note that kink or antikink obviously give an equal contribution.  Thus we write
\beq
\int dx \haf 1 \phi_1'^2=\int dx \haf 1 \phi_2'^2\equiv \haf {m_\phi}\quad{\rm and}\int dx \haf 1 \psi_1'^2=\int dx \haf 1 \psi_2'^2\equiv \haf {m_\psi}.\label{n}
\eeq
$m_\phi$ and $m_\psi$ should not be confused with the actual masses of the perturbative excitations of each respective field.  The contribution of the potential to the action will also be easily deconstructed.  There will be a contribution from around the region of each soliton and there will be a contribution from the region in between the solitons when they separate.  We find that the interaction potential plays no role in evaluating the action.  We will assume this for the moment, and discuss it in more detail in the next subsection.  
\subsection{Free kink approximation}
We can write
\beq
\int dx\, V(\phi,\psi)\approx \int_{x\approx x_1}dx\left(V_\phi+V_\psi\right)+\int_{x\approx x_2}dx\left(V_\phi+V_\psi\right)\label{t}
\eeq
and we will compute each term in Eqns.(\ref{n},\ref{t}) in the free kink approximation.  For a field $\chi$, which satisfies an equation with solution given by a free kink, we have a Lagrangian
\beq
L=\haf 1\chi'^2+V(\chi)\label{11}
\eeq
and equation of motion
\beq
\chi''-V'(\chi)=0.
\eeq
The first integral of the equation of motion is conservation of energy:
\beq
\haf 1\chi'^2-V(\chi)=C\to 0
\eeq
where $C$ is a constant which we normalize to zero.  Therefore $\chi'=\sqrt{2V(\chi)}$.  Then the integrals that appear in the action above are reduced to quadrature
\beq
\int_{x_i}^{x_f} dx \haf 1\chi'^2=\int_{x_i}^{x_f}  dx \haf 1\sqrt{2V(\chi)}\chi'=\int_{\chi_i}^{\chi_f}  d\chi \sqrt{V(\chi)\over 2}
\eeq
and
\beq
\int_{x_i}^{x_f} dx V(\chi)=\int_{x_i}^{x_f}  dx \haf 1\sqrt{2V(\chi)}\chi'=\int_{\chi_i}^{\chi_f}  d\chi \sqrt{V(\chi)\over 2};
\label{15}
\eeq
evidently, the integrals are equal.   
\subsection{Analytical evaluation of the Euclidean action}
For $\phi$, the potential is $V_\phi=\alpha\sin^2(\pi\phi)$ which gives
\beq
\haf{m_\phi}=\int_{x_i}^{x_f} dx \haf 1\phi'^2=\int_{x_i}^{x_f} dx V_\phi(\phi)=\sqrt{\alpha\over 2}\int_{-1}^{0}  d\phi (-\sin(\pi\phi))=\sqrt{\alpha\over 2}\left.\frac{\cos(\pi\phi)}{\pi}\right|^0_{-1}=\frac{\sqrt{2\alpha}}{\pi}.\label{massphi}
\eeq
The minus sign in front of the $\sin(\pi\phi)$ comes from the square root since for the first kink $\phi\in[-1, 0]$ where $\sin(\pi\phi)\le 0$.  For the field $\psi$ we have $V_\psi=\beta\left(\psi^2-a^2\right)^2$ which gives
\beq
\haf{m_\psi}=\int_{x_i}^{x_f} dx \haf 1\psi'^2=\int_{x_i}^{x_f} dx V_\psi(\psi)=\sqrt{\beta\over 2}\int_{-a}^{a}  d\psi (a^2-\psi^2)=\frac{2}{3}\sqrt{2\beta}a^3\label{masspsi}
\eeq
We note the sign change in the intermediate equation as $(\psi^2-a^2)$ is negative for $\psi\in[-a,a]$ and the limits $x_i$ and $x_f$ are assumed to be both close to $x_1$ or $x_2$ as required.

We will  evaluate the  energy, first in the metastable configuration and second in the configuration at the top of the potential barrier, so that eventually we can normalize the Euclidean action properly and then evaluate it.  For the metastable configuration, the energy essentially obtains contributions from the locations of the kink solitons, two for $\phi$ and two for $\psi$, hence
\bea
E_0&=&\haf 1\int dx  \left(\phi_1'^2+\phi_2'^2+\psi_1'^2+\psi_2'^2\right)+\int_{x\approx x_1\atop x\approx x_2} dx V(\phi,\psi)\cr
&=&(m_\phi+m_\psi)+\int_{x\approx x_1} dx \left(V(\phi)+V(\psi)\right)+\int_{x\approx x_2} dx (V(\phi)+V(\psi))\cr
&=&2(m_\phi+m_\psi)
\eea
using Eqns. (\ref{massphi},\ref{masspsi}).  We will subtract $E_0$ from the potential so that we normalize the initial configuration to vanishing energy.  

For the configuration at the top of the potential barrier we must take into account of the following differences:   the solitons must have kinetic energy in order to conserve energy,  we have two additional kinks in the $\psi$ field, and if the solitons separate by a distance $|x_2-x_1|$ the energy increases (it actually decreases as this contribution is negative) by $\int_{x_1}^{x_2} dxV(\phi,\psi)$.  For the metastable configuration we did not consider that the solitons could separate; it is assumed that they are herded together by the shepherd field.  If the solitons at the top of the potential barrier separate, they leave behind the true vacuum.  Since we will normalize the vacuum energy so that the false vacuum outside the shepherd soliton has zero energy density, the true vacuum between will have negative energy, so this contribution in fact decreases the energy.  The $\psi$ field is in its true vacuum outside the shepherd soliton; thus there is no energy density difference for it. However, the $\phi$ field is in its false vacuum outside the shepherd soliton; hence, looking at the potential Eqn. \eqref{Vphi}, we see that the contribution will be
\beq
\int_{x_1}^{x_2}dx\, V(\phi,\psi)=-\alpha\epsilon_\phi |x_2-x_1|.
\eeq
Then the energy for the configuration at the top of the barrier will be
\bea
E_T-E_0&=&\haf 1 (m_\phi+2m_\psi)(\dot x_1^2+\dot x_2^2) +2(m_\phi+2m_\psi) -2(m_\phi+m_\psi) -\alpha\epsilon_\phi |x_2-x_1|\cr
&=&\haf 1 (m_\phi+2m_\psi)(\dot x_1^2+\dot x_2^2) +2m_\psi -\alpha\epsilon_\phi |x_2-x_1|\cr
&=&T+V\label{energy}
\eea
The Minkowski action is $S_M=\int dt \,(T-V)$ while the energy is $E=T+V$.  Analytically continuing to Euclidean time, $T\to -T$ and  $S_M\to-\int d\tau\, ( T+V)=-S_E$.  Then the Euclidean action is $S_E=\int d\tau\, (T+V)$ and is easily read off as 
\beq
S_E=\int d\tau\, \left(\haf 1 (m_\phi+2m_\psi)(\dot x_1^2+\dot x_2^2) +2m_\psi -\alpha\epsilon_\phi |x_2-x_1|\right)
\eeq
where the overdot now means the derivative with respect to $\tau$.  The Euclidean trajectory, which we can  compute analytically, starts at the configuration at the top of the barrier. The solitons separate until they reach the bounce point, and then they come back together and coalesce.  We define center of mass and relative coordinates $X=(x_1+x_2)/2$ and $x=x_2-x_1$; in terms of these, $\dot x_1^2+\dot x_2^2=2\dot X^2+\haf 1 \dot x^2$.  We take the center of mass coordinate to be constant, $X=\dot X=0$ so that $\dot x_1^2+\dot x_2^2=\haf 1 \dot x^2$.  The Euclidean ``energy'', $T-V$,  is conserved and normalized to zero
\beq
0=T-V= \haf 1 \left(\haf{m_\phi}+m_\psi\right)\dot x^2 -2m_\psi +\alpha\epsilon_\phi x\label{ec}
\eeq
where we restrict $x>0$.  The solitons separate until $\dot x=0$, hence at that point $x=x_B$ which is given by
\beq
x_B=\frac{2m_\psi}{\alpha\epsilon_\phi}.
\eeq
Then the contribution to the action from this part of the trajectory will be
\beq
S_E=\int d\tau  \haf 1 \left(\haf{m_\phi}+m_\psi\right)\dot x^2 +2m_\psi -\alpha\epsilon_\phi |x| .\label{action}
\eeq
The energy condition Eqn. \eqref{ec} implies
\beq
\dot x=2\sqrt\frac{(2m_\psi-\alpha\epsilon_\phi x)}{{m_\phi}+2m_\psi}.
\eeq
Replacing, in the usual way as in Eqns. (\ref{11}-\ref{15}) for $\dot x$ in the first term of Eqn. \eqref{action} we find
\bea
S_E&=&\int d\tau  \haf 1 \left(\haf{m_\phi}+m_\psi\right)2\sqrt\frac{(2m_\psi-\alpha\epsilon_\phi x)}{{m_\phi}+2m_\psi}\dot x +\sqrt{2m_\psi -\alpha\epsilon_\phi x}\haf 1\sqrt{{m_\phi}+2m_\psi}\dot x\cr
&=&\int d\tau \sqrt{2m_\psi -\alpha\epsilon_\phi x}\sqrt{{m_\phi}+2m_\psi}\dot x\cr
&=&\int_{x_T}^{x_B} dx \sqrt{2m_\psi -\alpha\epsilon_\phi x}\sqrt{{m_\phi}+2m_\psi}=\left.\frac{-2\sqrt{{m_\phi}+2m_\psi}(2m_\psi -\alpha\epsilon_\phi x)^{3/2}}{3\alpha\epsilon_\phi} \right|^{x_B}_{x_T\approx 0}\cr
&=&\frac{2\sqrt{{m_\phi}+2m_\psi}(2m_\psi)^{3/2}}{3\alpha\epsilon_\phi}\label{evaction}
\eea
where $x_T$ is the value of $x$ at the top of the barrier, which is of the order of the size of the solitons, but we can approximate it as zero).  This is the contribution to the action of the part of the instanton from the top of the barrier to the bounce point. 

\subsection{Contribution of the interaction potential}
In this subsection we explain why the interaction potential $V_{\phi\psi}$ does not contribute significantly to the energy and the action.  If we imagine the sheep solitons separating, then between them the $\phi$ field will be in its true vacuum but the $\psi$ field will be in its false vacuum.  Thus as they separate, the total energy will grow linearly.  Thus the shepherd solitons give a compression that holds the sheep together.  However as the shepherd solitons compress the sheep more and more, there will be an increase in the pressure, and the system will come to an equilibrium.  There are many contributions to the pressure forces, the $\phi$ kinks becoming steeper than their natural slope will add energy and hence pressure.  But there will be a contribution coming from the interaction term.  Effectively the shepherd solitons would like to approach each other and annihilate completely in order to minimize their energy.  The sheep solitons get in the way.  The energy diminishes linearly as the separation of the shepherd solitons goes to zero, with coefficient $\beta \epsilon_\psi$.  If one forces the shepherd solitons closer and closer, the value of the $\psi$ field at the point where $\phi\approx1/2$, starts to move significantly away from $\pm a$, the values for which the numerator of the interaction term vanishes.  This will make the energy  increase, to first approximation linearly, however with much greater coefficient $\xi>\beta\epsilon_\psi$.  Thus the dynamics can be simply modelled as given by a potential of the form
\beq
V_{eff.}=\beta\epsilon_\psi x- \xi (x-x_0),
\eeq 
where the second term is valid only for $x\to x_0^-$ and $x_0$ is smaller than the size of the free solitons.  For smaller values of $x$ the repulsion becomes quite non-linear.  However, the energy in the compression term is comparable to $\beta\epsilon_\psi x_0$, which, and the corresponding pressure, for small $\epsilon_\psi$ can be made very small.  This compression will be balanced by an equal, and hence small, but opposing, repulsive force.  Thus  this contribution to the energy of the solitons is arbitrarily small in the limit $\epsilon_\psi\to 0$, which is what we assume.  Therefore, the interaction term does not contribute significantly to the energy of the solitons.  The most physical analogy can be made to nuclei, which have intrinsic mass energy, proportional to the number of nucleons, but which is modified by the binding energy of the nuclei.  For precision measurements the binding energies are important, but in the values of the overall masses, they are quite negligible. 

\subsection{Contribution of the trajectory from the initial metastable state to the top of the barrier}

\begin{figure}
\begin{center}
\includegraphics[width=7cm]{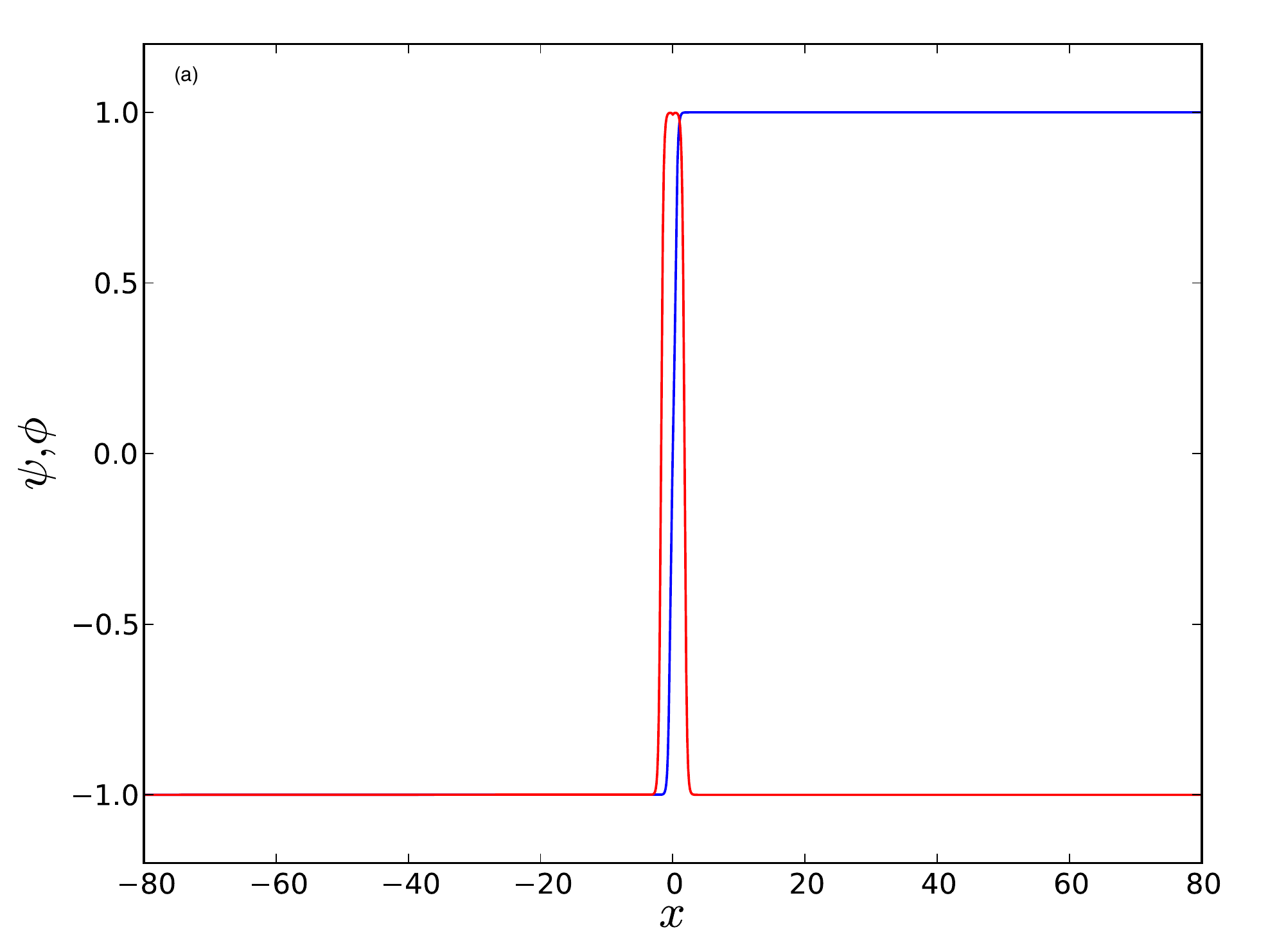}
\qquad
\includegraphics[width=7cm]{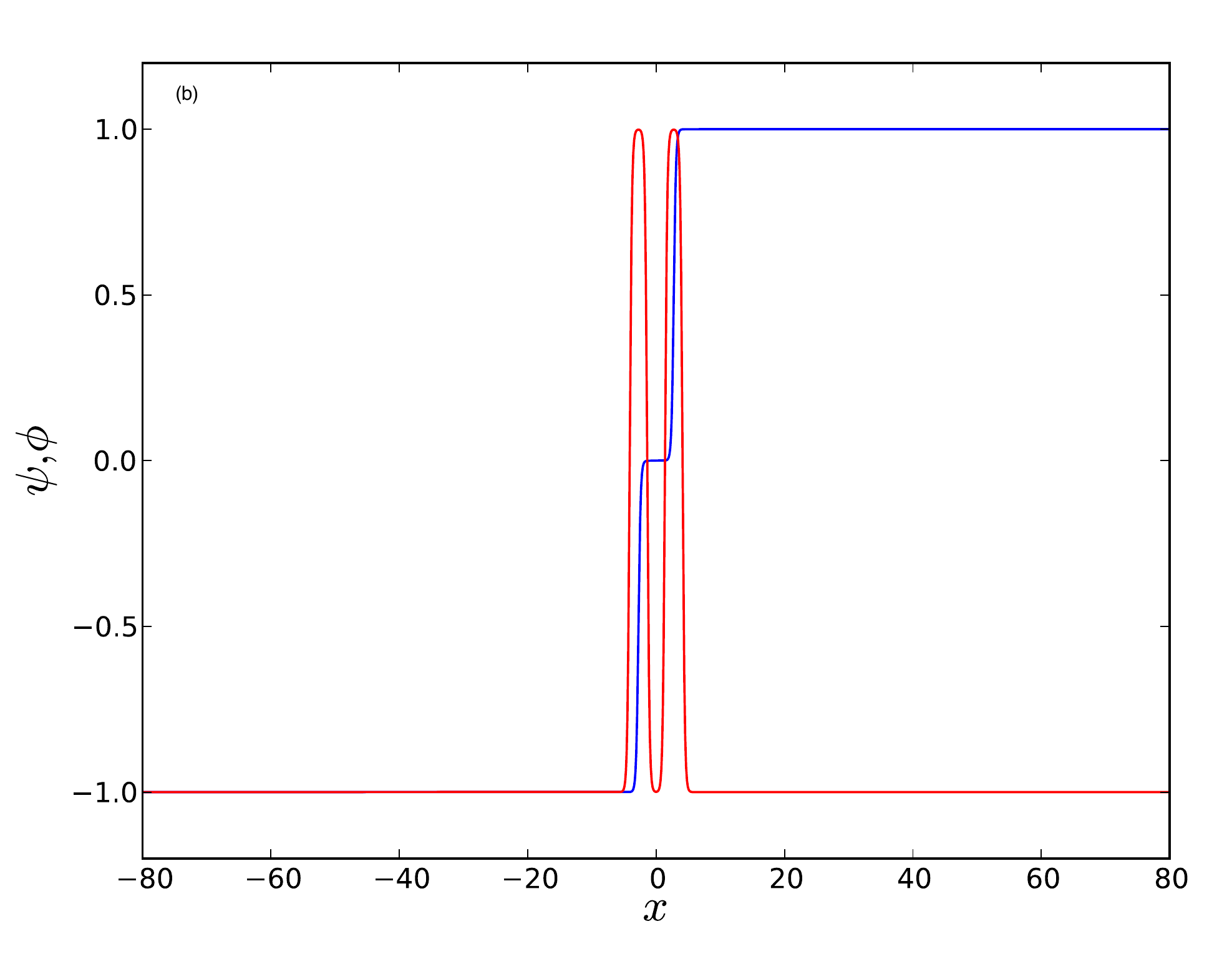}
\qquad
\includegraphics[width=7cm]{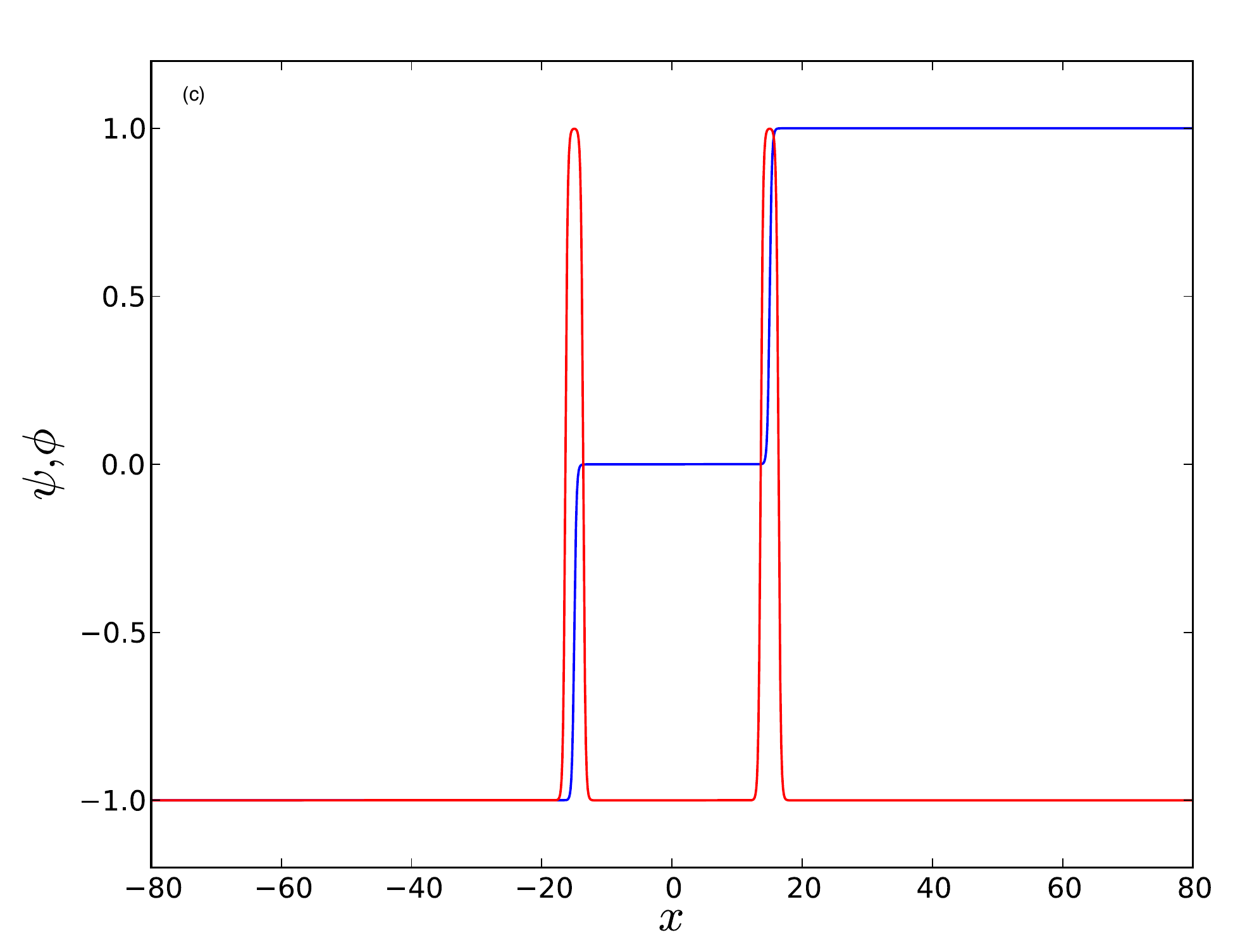}
\qquad
\includegraphics[width=7cm]{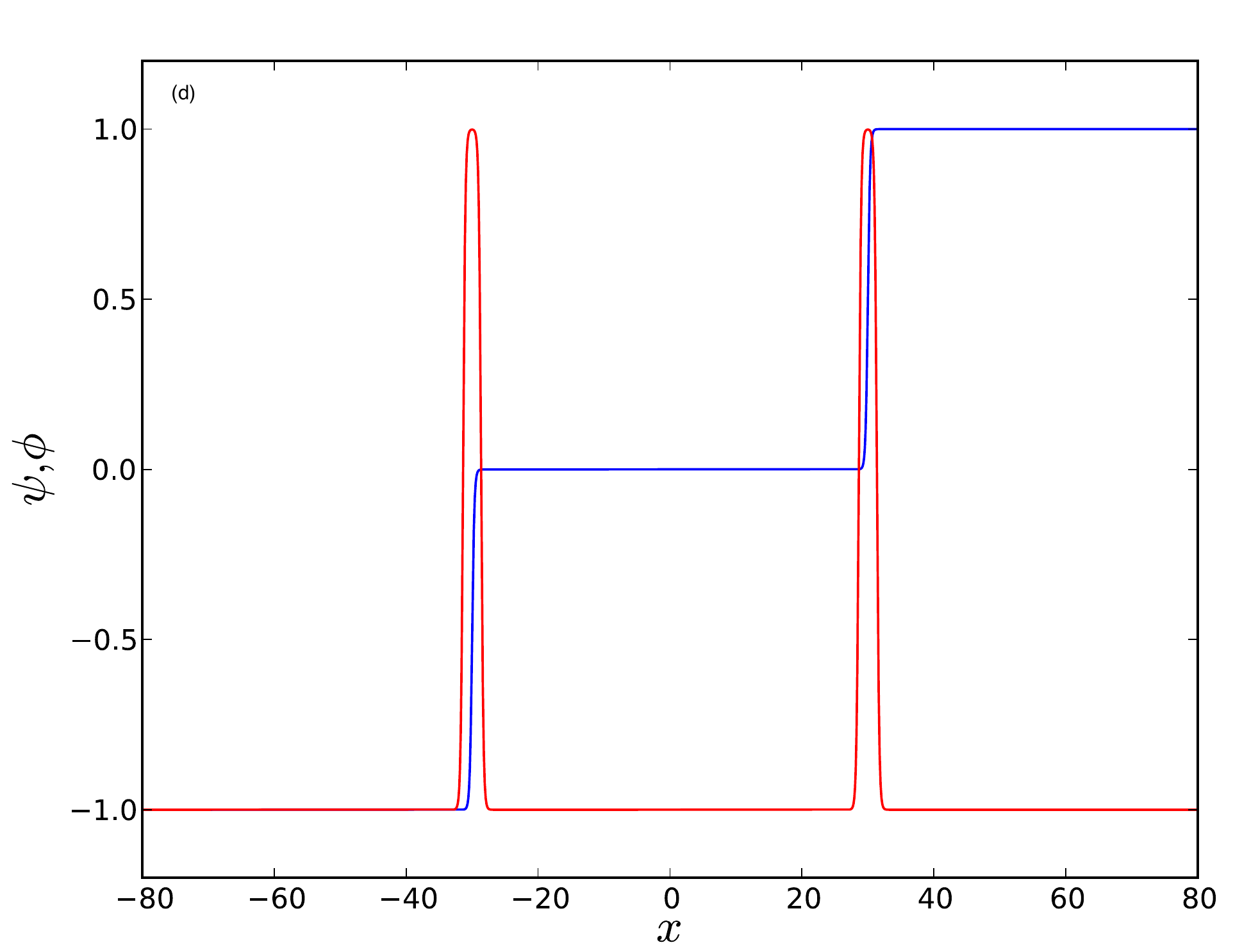}
\end{center}
\caption{(Color online) (a) Metastable soliton configuration. (b) Solitons just touching. (c) Solitons separated by a macroscopic amount. (d) Solitons separated by a macroscopic amount. } 
\label{fig6}
\end{figure}

The evolution from the initial metastable configuration to the top of the potential barrier, involves passage through configurations that cannot be well approximated by the free kinks.  We would have to resort to numerical methods to calculate the contribution to the action of that part of the trajectory, however, we will in fact be able to conclude that it is negligible.  With a simple parametrization of the configuration in terms of hyperbolic tangents, we can get an estimate of the contribution to the action from the initial part of the trajectory.  Taking
\bea
\psi(x)&=&\haf a\left(\tanh \Omega(x+(\xi/2+(2/\omega)))-\tanh \Omega(x+(\xi/2-(2/\omega)))\right.\cr
&+&\left.\tanh\Omega(x-(\xi/2-2/\omega))-\tanh\Omega(x-(\xi/2+2/\omega))-1\right)\cr
\phi(x)&=&\haf 1\left(\tanh\omega(x+\xi/2)+\tanh\omega(x-\xi/2)\right),
\eea
this function is depicted in Fig. (\ref{fig6}a,b).  

\begin{figure}
\begin{center}
\includegraphics[width=\textwidth]{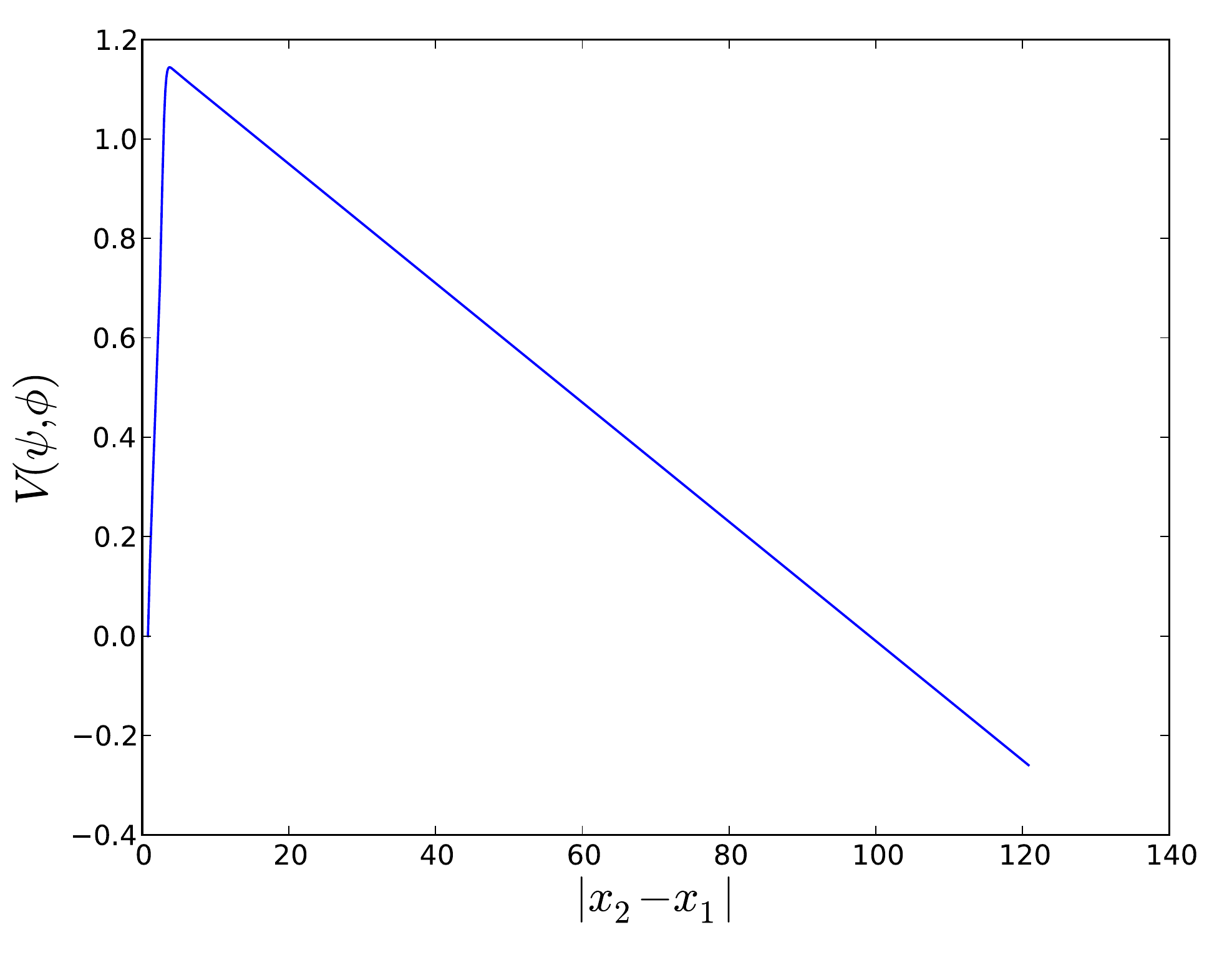}
\end{center}
\caption{(Color online) Potential for the parametrization of the bounce trajectory. } 
\label{fig7}
\end{figure}

\begin{figure}
\begin{center}
\includegraphics[width=\textwidth]{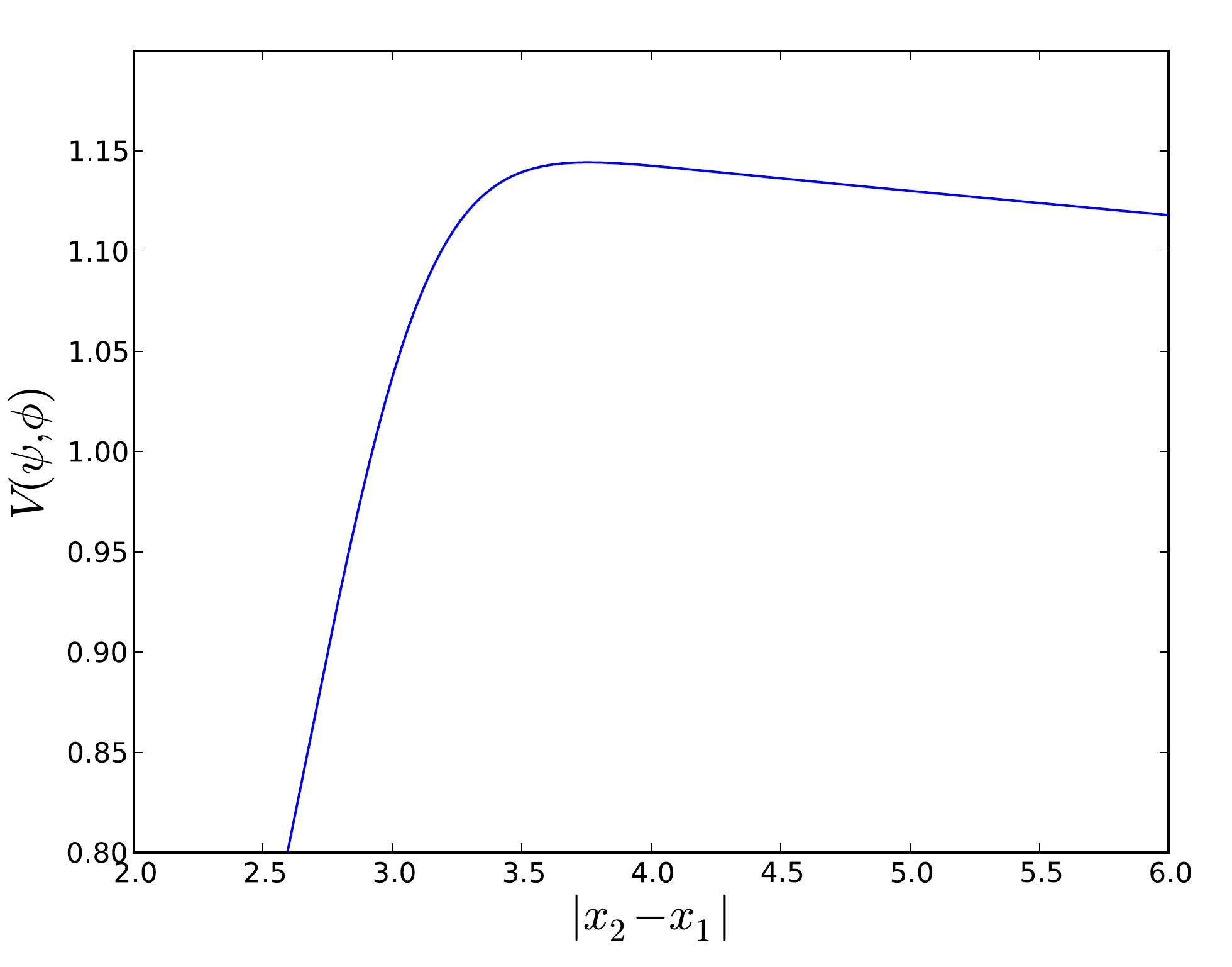}
\end{center}
\caption{(Color online) Zoom of the top of the barrier of the potential for the parametrization of the bounce trajectory. } 
\label{fig8}
\end{figure}

The numerically evaluated action for this trajectory gives an upper bound for the contribution to the action from the part of the trajectory  which corresponds to the formation of the solitons at the top of the barrier from the initial metastable configuration.  The subsequent contribution to the action has already been evaluated in the previous section in Eqn. \eqref{evaction}.
The numerical evaluation of the action is shown in Figure \eqref{fig7}, with a zoom of the top of the barrier in Figure \eqref{fig8} in order to be confident that it is smooth. 
This behaviour of the  potential  is completely understandable in terms of the energy found in Eqn. \eqref{energy}, which corresponds to the part of the potential starting at the top of the barrier and descending with the very slightly sloped straight line.  The bounce point is achieved at $x_B={2m_\psi}/{\alpha\epsilon_\phi}$, which obviously behaves as $o(1/\epsilon_\phi)$.  The trajectory from the metastable state to the top of the barrier, in the parametrized set of configurations, achieves the top of the barrier with a variation of $x$ which is of the order of the size of the solitons, a size which is $o(1)$ when compared with $o(1/\epsilon_\phi)$.  Using an analysis identical to that preceding Eqn. \eqref{evaction}, the contribution to the Euclidean action, with a linear approximation to the potential, is given by
\beq
S_E=\int^{x_T}_{0} dx \sqrt{2m_\psi x}\sqrt{{m_\phi}+2m_\psi}= \sqrt{2m_\psi }\sqrt{{m_\phi}+2m_\psi}{x_T}^{3/2}\sim o(1).
\eeq
Thus it is perfectly reasonable to neglect this contribution to the Euclidean action.

\section{Discussion}\label{sec-discussion}
We have computed the decay of a domain wall which traps the true vacuum inside the false vacuum.  At first glance, one would think that such a domain wall would be classically unstable to dissociation, however it is easy to conceive of interactions which keep the soliton classically stable.  The model presented here may be artificial; however, the general idea is clear.  Any potential which prevents the passage of the sheep solitons through the shepherd solitons will do.

There is furthermore no analogous surface energy that occurs in two \cite{vortex,bulge} and three \cite{kpy} spatial dimensional examples,  which could trap the true vacuum inside, at least classically.  However, in this paper we find that we can build on the analogy with the two spatial dimensional models \cite{vortex,bulge}.  In those models, the true vacuum was separated from the false vacuum by a thin wall.  The thin wall was achieved by having several topological quanta of one of the fields trapped inside the wall.  The pressure of these quanta made the wall thin.  It may well be that the thin wall monopole suggested in \cite{kpy} in fact only occurs for large monopole charge as has been found in several large charged BPS monopoles \cite{bm}.

In the present paper, we also trap $N$ solitons of the field $\phi$ inside the confining soliton of the field $\psi$.  The $N$ solitons in principle exert a pressure and the overall size of the full soliton is proportional to $N$.  The decay of the soliton can occur through tunnelling transitions.  In the initial metastable configuration, the external false vacuum is due to the $\phi$ field; the $\psi$ field is in its true vacuum.  However, inside, the $\psi$ field goes to its false vacuum while the $\phi$ field interpolates from the false vacuum to the true vacuum and back to the false vacuum multiple times, finally exiting the confining region in its false vacuum.   Tunnelling transitions, essentially in the field $\psi$ inside the confining soliton, will  cut the soliton into two parts.   The region between the two parts is in the true vacuum for both fields.  This could happen at any point within the confining soliton where the $\phi$ field  crosses its true vacuum, which occurs multiple times if there are multiple $\phi$ solitons confined.   It will continue to occur until all the $\phi$ solitons are free.  The configuration will then resemble one half-soliton, linking the false vacuum on the outside to the true vacuum, followed by a string $N$ of full solitons, in which the $\phi$ field makes the transit from true to adjacent false to the subsequent true vacua while the $\psi$ field makes the transition true to false and back to true vacua, followed by a final half-soliton interpolating from the true vacuum to the false vacuum on the outside at the other end.  However, after any one transition, the false vacuum outside will be unstable to the groups of herded solitons from separating apart, converting false vacuum to true vacuum.  The decay rate for each tunnelling transition is proportional to
\beq
\Gamma\sim e^{-{2\sqrt{{m_\phi}+2m_\psi}(2m_\psi)^{3/2}}/{3\alpha\epsilon_\phi}}\kappa
\eeq
where $\kappa$ is the usual determinant prefactor \cite{Coleman, calcol77}.  We note that parametrically for $\alpha$ large enough, this decay rate can be unsuppressed.

\section*{Acknowledgements}
We thank NSERC of Canada for financial support.  We thank St. John's College and DAMTP, both of Cambridge University for hospitality, where this work was completed.  Some of the work of MH was undertaken at DAMTP University of Cambridge, financially supported by the UK Science and Technology Facilities Council under grant number ST/J000434/1.  We thank Nick Manton for useful discussions.

\end{document}